\documentclass[prl,twoside,twocolumn,showpacs]{revtex4}

\usepackage{amsmath}
\usepackage{amssymb}
\usepackage{latexsym}
\usepackage{amsfonts}
\usepackage{graphicx}

\newcommand{\bra}[1]{\langle #1 |}
\newcommand{\ket}[1]{| #1 \rangle}

\newcommand{\matx}[1]{| #1 \rangle \langle #1 |}

\begin{document}
\title{Composite Geometric Phase for Multipartite Entangled States}
\author{M. S. Williamson}
\author{V. Vedral}
\affiliation{School of Physics and Astronomy, University of Leeds,
Leeds, LS2 9JT, UK.}
\date{\today}

\begin{abstract}
When an entangled state evolves under local unitaries, the
entanglement in the state remains fixed. Here we show the dynamical
phase acquired by an entangled state in such a scenario can always
be understood as the sum of the dynamical phases of its subsystems.
In contrast, the equivalent statement for the geometric phase is not
generally true unless the state is separable. For an entangled state
an additional term is present, the mutual geometric phase, that
measures the change the additional correlations present in the
entangled state make to the geometry of the state space. For $N$
qubit states we find this change can be explained solely by
classical correlations for states with a Schmidt decomposition and
solely by quantum correlations for W states.
\end{abstract}
 \maketitle

In a seminal paper \cite{ref:Berry84}, Berry recognized that a
quantum system undergoing a cyclic, adiabatic evolution records the
path of its evolution as a geometrical quantity in the phase of its
wavefunction. This phase, the geometric phase (GP), forms part of the
total phase of wavefunction in addition to the more familiar
dynamical phase. Since Berry's initial discovery, the
GP has been found to occur in more general
circumstances; non-adiabatic \cite{ref:Aharonov&Anandan87} and
non-cyclic evolutions \cite{ref:Samuel&Bhandari88}, non-abelian form
\cite{ref:Wilczek&Zee84} and for states that are mixed
\cite{ref:Sjoqvist00}.

GPs are of interest for many reasons, amongst them topological
effects in many body systems \cite{ref:ShapereBook} and their use
for quantum information processing \cite{ref:Jones99}, a paradigm
where entanglement is known to be key in its advantage over the
classical counterpart. GPs may also be used as a tool. Because they
depend on the geometry of the space the states traverse, GPs can
provide information on this space. As an example, this property can
be used to discover the coordinates of quantum phase transitions in
a parameter space \cite{ref:Carollo&Pachos05}. On another currently
popular front, entanglement has been shown to be present in many
body systems at finite temperatures \cite{ref:Ghosh03}. Studying GPs
of multipartite entangled states may therefore prove to be
complimentary to both efforts.

GPs have been studied only for entangled bipartite systems so far.
These include qubits precessing in magnetic fields
\cite{ref:Sjoqvist00a} and general evolution \cite{ref:Tong03b} both
without interaction (fixed entanglement) and various specific
Hamiltonians for bipartite systems with interactions (changing
entanglement) \cite{ref:Fuentes-Guridi02,ref:Yi04a}. It has been
shown \cite{ref:Sjoqvist00a,ref:Tong03b} in bipartite systems that
even if there are no interactions during the evolution, fixed
entanglement affects the geometric phase.

It is this last relation that we investigate further in this Letter
for multipartite systems. Of the two terms making up the total phase
of a quantum state, dynamical and geometric phase, we will show that
under local evolutions the dynamical phase of a composite state can
always be understood as the sum of the dynamical phases that
comprise the composite system whether the state is entangled or not.
We will also show that the same statement cannot be made for GP when
the state is entangled. When the composite state is entangled an
additional correction term, much like an interference term, $\Delta
\gamma$, is present. Explicitly $\Gamma=\sum_{n=1}^N \gamma_n^M
+\Delta \gamma$ where the composite state GP is $\Gamma$ and the
$n^{th}$ of the $N$ subsystems has GP $\gamma_n^M$. The correction
term we call the mutual GP. Each subsystem, $\rho_n$, is obtained by
tracing over the other $N-1$ subsystems in the composite state
$\rho$.

In tracing out the subsystems from the composite state we have
removed both quantum and classical correlations
\cite{ref:Henderson&Vedral01}. The extra correlations present in the
composite state, unaccessible when one only has access or control on
one of the subsystems, modify the GP but not the dynamical phase
when the state is entangled. Should one only have access or control
on one of the subsystems, $\rho_n$, then a GP of $\gamma_n^M$ would
be observed. As previously noted GPs are dependent on the underlying
geometry of the quantum state space and here we are using the GP as
a tool to sample the change these correlations have on the geometry
of the state space. We characterize which type of correlations,
quantum (entanglement) or classical, are responsible for the
correction, $\Delta \gamma$, for two specific types of state we know
how to quantify the entanglement of using the relative entropy of
entanglement, $E_R$ \cite{ref:Vedral97}. We will show $\Delta
\gamma$ can be attributed solely to classical correlations for
states with a Schmidt decomposition (labeled hereafter as S states)
and solely to entanglement for W states.

Our aim is to study the effect of fixed entanglement on the mutual
GP, $\Delta \gamma$. We fix entanglement during the evolution by
requiring that the evolution on the composite state, the unitary
$\mathcal{U}$, be composed of local unitaries, $U_n$, acting on each
subsystem, $n$. We write this as
\begin{equation}\label{eq:compositeunitary}
\mathcal{U}(t)=\bigotimes_{n=1}^N U_n(t).
\end{equation}
With this condition we can show that the dynamical phase of the
composite system, $\Delta$, is always given by the sum of the
dynamical phases of its subsystems, $\delta_n^M$. The definition of
dynamical phase \cite{ref:Sjoqvist00} is
\begin{equation}\label{eq:dynamicalphaseformula}
\Delta=-i\int_0^T tr[\rho\mathcal{U}^\dag(t) \dot{\mathcal{U}(t)}]dt.
\end{equation}
Substituting eq~(\ref{eq:compositeunitary}) into this definition we
see
\begin{eqnarray}
\Delta=&-i\int_0^T tr[\rho\sum_{n=1}^N U_n^\dag(t) \dot{U}_n(t)
\bigotimes_{m=1,m\neq n}^N \mathbb{I}_m]dt \nonumber \\
=&\sum_{n=1}^N-i\int_0^T tr[\rho_n U_n^\dag(t) \dot{U}_n(t)]dt,
\end{eqnarray}
showing $\Delta=\sum_{n=1}^N \delta_n^M$. The composite dynamical
phase is always equal to the sum of the subsystem dynamical phases
whether the state is entangled or not. This statement holds because
we have constrained the dynamics to be local. Before demonstrating
this is not the case for GP we discuss parallel transport as a
useful way to isolate the GP.

A quantum state is said to be parallel transported when it acquires
no dynamical phase at each point along its evolution. Formally the
mathematical condition that requires the state is in phase with
itself at each point is
$tr\rho\mathcal{U}(t)^\dag\dot{\mathcal{U}}(t)=0$ $\forall$ $t$. If
the state is parallel transported (by $\mathcal{U}^\|$) then the
total phase obtained by the state will be equal to the GP.
Pancharatnam \cite{ref:Pancharatnam56} gave a natural prescription
to obtain the total phase, generalized to mixed states
\cite{ref:Sjoqvist00}. As mentioned, under parallel transport
conditions this becomes the GP, $\Gamma$, accumulated over the time
$t\in[0,T]$.
\begin{equation}\label{eq:geometricphaseformula}
\Gamma=\arg\{tr\rho\mathcal{U}^\|(T)\}.
\end{equation}
For unentangled states of the form $\rho=\bigotimes_{n=1}^N \rho_n$
we see that under local unitaries
eq~(\ref{eq:geometricphaseformula}) becomes
$\Gamma=\arg\left\{\prod_{n=1}^N tr\rho_n U_n^\|(T)\right\}$ so
$\Gamma=\sum_{n=1}^N \gamma_n^M$. But in general we see that
$\Gamma\neq\sum_{n=1}^N \gamma_n^M$.

It is known that there are many parallel transport conditions for
mixed states. In this study the subsystems, $\rho_n$, will be the
mixed states. Some of these parallel transport conditions produce
GPs that are a property of an arbitrary entangled purification of
the mixed state in a higher dimensional space, not just of the mixed
state itself. It is also known, however, that a subset of these
conditions, those proposed by \cite{ref:Sjoqvist00} (`stronger'
parallel transport conditions) provide a mixed state GP that is a
property of the evolution of the mixed state only
\cite{ref:Ericsson03b}. For this reason we use the stronger parallel
transport conditions to constrain the local unitaries. These
conditions require that every eigenvector, $\ket{\phi_i^n}$, of each
subsystem, $n$, be parallel transported, formally
$\bra{\phi_i^n}U_n^\dag(t) \dot{U}_n(t)\ket{\phi_i^n}=0$ $\forall$
$i,n,t$. We write the unitary acting on subsystem $n$ that fulfills
these conditions $U_n^\|$.

We now calculate $\Gamma$ and $\gamma_n^M$ for arbitrary
superpositions of W states which contain, amongst others, W, S and
GHZ states. We will write these states as
\begin{equation}\label{eq:superpositionsymmetricstates}
\ket{\Psi(0)}=\sum_{k=0}^N a_k \ket{N,k},
\end{equation}
where $a_k$ are the complex probability amplitudes and the W state
$\ket{N,k}$ is defined by
\begin{equation}
\ket{N,k}=\frac{1}{\sqrt{{\tiny\left(
                           \begin{array}{c}
                             N \\
                             k \\
                           \end{array}
                         \right)}}}\hat{S}\ket{\underbrace{000}_{N-k}....\underbrace{111}_k}.
\end{equation}
$\hat{S}$ is the total symmetrization operator. Time evolution of
the state $\ket{\Psi(0)}$ is given by
$\ket{\Psi(t)}=\mathcal{U}(t)\ket{\Psi(0)}$. Each individual qubit
in the state considered on its own is given by the density matrix
$\rho_n(t)=tr_{1..N,\neq n}\ket{\Psi(t)}\bra{\Psi(t)}$.
In the $\ket{0}$, $\ket{1}$ basis the subsystem state explicitly is
\begin{equation}
\rho_n(0)=\rho_{00}\matx{0}+\rho_{11}\matx{1}+\rho_{01}\ket{0}\bra{1}+\rho_{01}^*\ket{1}\bra{0},
\end{equation}
where $\rho_{00}=\sum_{k=0}^N |a_k|^2 \frac{N-k}{N}$,
$\rho_{11}=\sum_{k=0}^N |a_k|^2 \frac{k}{N}$ and
$\rho_{01}=\sum_{k=1}^N a_{k-1}^* a_k \sqrt{k(N-k+1)}/N$. The GP for
both the composite state $\ket{\Psi(T)}$ and each of its subsystems,
$\rho_n(T)$, is made by substitution into
eq~(\ref{eq:geometricphaseformula}). We have
$\Gamma=\arg\{\sum_{k,l=0}^N a_l^* a_k
\bra{N,l}\mathcal{U}^\|(T)\ket{N,k}\}$. Cross terms disappear from
this equation when each local unitary, $U_n^\|$, brings each
subsystem back to the same ray ($U_n^\|(T)\ket{0}=U_n^\|(0)\ket{0}$
up to a global phase). Under this evolution the global GP, $\Gamma$,
becomes
\begin{equation}\label{eq:gpcomposite}
\Gamma=\arg\left\{\sum_{k=0}^N \frac{|a_k|^2}{\tiny{\left(
                                         \begin{array}{c}
                                           N \\
                                           k \\
                                         \end{array}
                                       \right)}} \sum_{m=1}^{\tiny{\left(
                                         \begin{array}{c}
                                           N \\
                                           k \\
                                         \end{array}
                                       \right)}} e^{i\sum_{n=1}^N
                                       A_{mn}^k\gamma_n}\right\},
\end{equation}
expressing the composite system GP in terms of pure state qubit
phases, $\gamma_n$. We have used the polar representation to define
$c_n e^{i\gamma_n}:=\bra{0}U_n(T)\ket{0}$ and its complex conjugate
$c_n e^{-i\gamma_n}:=\bra{1}U_n(T)\ket{1}$. $\gamma_n$ ($-\gamma_n$)
are the GP that the pure qubit states $\ket{0}$ ($\ket{1}$) acquire
over the evolution $U_n(T)$ and $c_n=1$ when each subsystem unitary
is cyclic. The matrix $A^k$ has elements $1(-1)$ to capture the sign
of $\gamma_n$ ($-\gamma_n$) in the exponent. Each ordered row in
$A^k$ has $N-k$ elements with value $1$ and $k$ elements of $-1$.
There are ${\tiny{\left(
                                         \begin{array}{c}
                                           N \\
                                           k \\
                                         \end{array}
                                       \right)}}$ ordered rows in $A^k$
                                       given by
all permutations of the first row elements.

Eq~(\ref{eq:gpcomposite}) can be understood as follows: Each
W state, $\ket{N,k}$, is a superposition of ${\tiny{\left(
                                         \begin{array}{c}
                                           N \\
                                           k \\
                                         \end{array}
                                       \right)}}$ kets, each with phase
factor $e^{i\sum_{n=1}^N A_{mn}^k\gamma_n}$. The phase factors for
each ket are averaged in the $m$ summation resulting in an overall
phase factor for each W state ($k$). The phase factors for
each W state are then averaged (the $k$ summation) to give
the total overall GP factor of the state.

The states of the subsystems are given by $\rho_n$ and the GP
associated to the them is also obtained by substitution into
eq~(\ref{eq:geometricphaseformula}) with $U_n^\|(T)$. We
substitute the general form of the eigenvectors and eigenvalues in
to this equation to obtain
$\gamma^M_n=\arg\{\frac{1}{2}(1+r)\bra{\phi_1^n}U^\|_n(T)\ket{\phi_1^n}
+\frac{1}{2}(1-r)\bra{\phi_2^n}U^\|_n(T)\ket{\phi_2^n}\}$, where the
general eigenvectors of a two level system are
$\ket{\phi_1^n}=e^{\frac{-i\phi_n}{2}}\cos(\theta_n/2)\ket{0}+e^{\frac{i\phi_n}{2}}\sin(\theta_n/2)\ket{1}$
and
$\ket{\phi_2^n}=-e^{\frac{-i\phi_n}{2}}\sin(\theta_n/2)\ket{0}+e^{\frac{i\phi_n}{2}}\cos(\theta_n/2)\ket{1}$.
Rewriting in the $\ket{0}$, $\ket{1}$ basis and substitution of
$\gamma_n$ identities results in the cyclic GPs of the
subsystems to be $\gamma_n^M=\arg\{\cos\gamma_n+ir\cos\theta_n\sin\gamma_n\}$. This can also be written as
\begin{equation}\label{eq:gpsubsystem}
\gamma_n^M=\arctan\left\{ \left( \sum_{k=0}^N |a_k|^2
\frac{N-2k}{N}\right)\tan\gamma_n \right\}
\end{equation}
for non-degenerate eigenvalues of $\rho_n$. We can interpret this
result in a similar manner to the pure state result. Writing
eq~(\ref{eq:gpsubsystem}) as $\gamma_n^M=\arg\left\{\sum_{k=0}^N |a_k|^2 \frac{N-k}{N}
e^{i\gamma_n}+\sum_{k=0}^N |a_k|^2 \frac{k}{N}
e^{-i\gamma_n}\right\}$, we can see $\gamma_n^M$ is an average of the phase factors of the
states $\ket{0}$ and $\ket{1}$ weighted by their relative frequency
in the composite state. For the remainder of this Letter we will
discuss specific cases of eqs~(\ref{eq:gpcomposite}) and
(\ref{eq:gpsubsystem}).

To simplify things, each local unitary will now be identical i.e.
$\mathcal{U}^\|= \bigotimes_{n=1}^N U^\|$ in all the following. All
pure state qubit GPs, $\gamma_n$, also become identical and will be
labeled $\gamma$.

The first specific example is that of S states. For these states
only two amplitudes in eq~(\ref{eq:superpositionsymmetricstates})
are non-zero, $a_0$ and $a_N$. The S state is
\begin{equation}
\ket{\Psi(0)}=\sqrt{\frac{1}{2}(1+r)}\ket{0}^{\otimes
N}+\sqrt{\frac{1}{2}(1-r)}\ket{1}^{\otimes N}.
\end{equation}
We have written $a_0$ and $a_N$ as the square root of the
eigenvalues of a bipartite state. Substitution into
eq~(\ref{eq:gpcomposite}) yields
\begin{eqnarray}
\Gamma=&\arg\{\frac{1}{2}(1+r)e^{iN\gamma}+\frac{1}{2}(1-r)e^{-iN\gamma}\} \nonumber \\
=&\arctan\{r\tan N\gamma \}.
\end{eqnarray}
In this formula pure state phases rather than phase factors add up.
Each of the local GPs for the S state are
\begin{equation}\label{eq:mixedcatgp}
\gamma^M=\arctan\{r\tan\gamma\}.
\end{equation}
To quantify the amount of entanglement in the composite states we
use the relative entropy of entanglement, $E_R$
\cite{ref:Vedral97,ref:Vedral&Plenio98}. For S states $E_R$ is given
by
\begin{equation}
E_R=1-\frac{1}{2}[(1+r)\log_2(1+r)+(1-r)\log_2(1-r)].
\end{equation}
$E_R$ takes a maximum value of $1$ when $r=0$ and a minimum value
$0$ when $r=1$. When a S state is maximally entangled ($r=0$)
$\Gamma$ can only take the values $0$ and $\pi$, however
eq~(\ref{eq:mixedcatgp}) for $\gamma^M$ is no longer valid as
$\rho_n$ is degenerate and must be evaluated via path ordering (see
\cite{ref:Singh03}). At the other extreme, separable states ($r=1$),
we see $\Gamma=N\gamma^M$ as expected.

The second specific example is a W state
$\ket{\Psi(0)}=\ket{N,k}$. The identical local evolution in this
case is particularly appropriate as these states often occur when
subsystems are indistinguishable. For this state $|a_k|^2=1$ and the
composite GP takes a particularly simple form
\begin{equation}
\Gamma=(N-2k)\gamma.
\end{equation}
For a W state the GP is the sum of the pure
state qubit phases much like a pure separable state. This result in
fact holds for any arbitrary superposition of W kets (for
example $a_1\ket{001}+a_2\ket{010}+a_3\ket{100}$). However the
subsystems are not generally pure and their GPs are
generally not $\gamma$ and $-\gamma$. They are
\begin{equation}\label{eq:mixedsymgp}
\gamma^M=\arctan\left\{ \left(\frac{N-2k}{N}\right)\tan\gamma
\right\}.
\end{equation}
The relative entropy of entanglement for W states is known
to be \cite{ref:Wei04,ref:Vedral04}
\begin{equation}
E_R=-\log_2\left(
               \begin{array}{c}
                 N \\
                 k \\
               \end{array}
             \right) - (N-k)\log_2\left(\frac{N-k}{N}\right) -
             k\log_2\left(\frac{k}{N}\right).
\end{equation}
Entanglement is maximal when $N=2k$ in which case
$E_R=N-\log_2{\tiny\left(
               \begin{array}{c}
                 N \\
                 N/2 \\
               \end{array}
             \right)}$ and is minimal ($E_R=0$) when $k=0,N$ (the
             state is separable). When the W state is
             maximally entangled $\Gamma=0$. This limit is similar to the maximally entangled
             S state, except it could also take the value
             $\pi$. We also note, for the same reason as maximally entangled S states,
             eq~(\ref{eq:mixedsymgp}) is not valid for calculation
             of $\gamma^M$ when $N=2k$. Again $\Gamma=N\gamma^M$
             for separable states (when $k=0,N$).

We now have the ingredients to calculate mutual GP, $\Delta \gamma$,
written explicitly as
\begin{equation}
\Delta \gamma = \Gamma - \sum_{n=1}^N \gamma_n^M.
\end{equation}
As previously mentioned, we have removed the quantum (entanglement)
but also the classical correlations in tracing out each subsystem
from the composite. We can characterize which missing correlations
are responsible for the correction term $\Delta \gamma$ using ideas
from entanglement distance measures, specifically here, the relative
entropy of entanglement, $E_R$.

To calculate $E_R$, the closest separable state, $\rho_S$, to the
entangled state needs to be found. Here closest means the separable
state from the set of all separable states that minimizes the
quantum relative entropy $S(\rho||\rho_S)=tr(\rho \log \rho - \rho
\log \rho_S)$, between it and the entangled state, $\rho$. The
minimum value of $S(\rho||\rho_S)$ is equal to $E_R$. The procedure
maximizes classical correlations between the entangled and the
separable state, any correlations left over are quantum. The closest
separable states are given in \cite{ref:Vedral&Plenio98} for S
states and \cite{ref:Wei04,ref:Vedral04} for W states.
The GPs, $\Gamma_S$, for these closest separable states are
\begin{equation}\label{eq:closestsepcat}
\Gamma_S=\arg\{\frac{1}{2}(1+r)e^{iN\gamma}+\frac{1}{2}(1-r)e^{-iN\gamma}\}
\end{equation}
for the S state and
\begin{equation}\label{eq:closestsepsym}
\Gamma_S=N\arctan\left\{\left(\frac{N-2k}{N}\right)\tan\gamma\right\}
\end{equation}
for the W state. In the latter we have used the binomial theorem and
the property $N\arg\{a\}=\arg\{a^N\}$.

\begin{figure}
\begin{center}
\includegraphics[width=8.6cm]{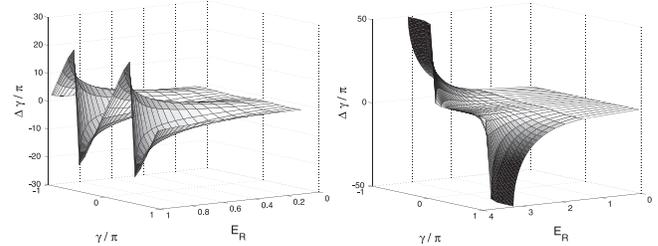}
\end{center}
\caption{$\Delta \gamma$ plotted against entanglement, $E_R$, and
$\gamma$ for $N=51$. Left panel: S state (classical correlations
only). Right panel: W state (quantum correlations only). $\Delta
\gamma$ is non-modular and plotted in the large $N$ limit with
varying path, $\gamma$, and entanglement in the composite state,
$E_R$. For constant $\gamma$ S states are monotonic w.r.t. $E_R$. W
states are monotonic except for the region $|\gamma|/\pi<1/2$.
Periodicities w.r.t. $\gamma$ are $\pi$ and $2\pi$ for S and W
states respectively when $E_R$ is fixed. For both states $N$
controls the magnitude of
$\Delta\gamma$.}\label{fig:Deltagammacombo}
\end{figure}

By taking the difference between the GPs of the composite state and
the closest separable state ($\Gamma-\Gamma_S$) we can see
exclusively what difference the quantum correlations make to the GP.
For S states this difference is zero as $\Gamma=\Gamma_S$. We can
therefore state quantum correlations have no effect on GP and state
space geometry. We can also state the correction term $\Delta
\gamma$ is the sole result of classical correlations for S states.
In contrast, $\Gamma-\Gamma_S=\Delta\gamma$ for W states because
$\Gamma_S=\sum_{n=1}^N \gamma^M$. Here we can state $\Delta \gamma$
results solely from quantum correlations. One common feature the
states share is that entanglement does not affect curvature when
$E_R \leq 1$. Note however that S states can have a continuum of
values $0 \leq E_R \leq 1$ independent of $N$. $E_R$ of W states
does have $N$ dependance but discrete values. The smallest of these
values are $E_R=0$ (separable, $k=0,N$), $E_R=1$ (singlet state,
$N=2,k=1$) or $E_R\geq \log_2 e$ for $k=1,N-1$. According to the
relative entropy of entanglement, perhaps counter intuitively, W
states are more quantum than S states, in the sense that classical
correlations cannot approximate these states as closely. This is
related to the robustness of the entanglement. Almost all
correlations in S states are classical, tracing out just one
subsystem results in a separable state. In contrast W states remain
entangled to the last pair of subsystems. Plotted in figure
\ref{fig:Deltagammacombo} is $\Delta \gamma$ for S and W states, the
difference classical and quantum correlations make to the GP
respectively. Although we don't know how to quantify the
entanglement of states like
eq~(\ref{eq:superpositionsymmetricstates}) we expect that $\Delta
\gamma$ for these states can be attributed to a mixture of classical
and quantum correlations, each reducing to what are probably the two
special limiting cases, S and W states.

In this Letter we have demonstrated that under local evolutions,
entangled states gain a correction term to their GP not present in
the dynamical phase. This correction term $\Delta \gamma$ has its
origins in the change the extra correlations present in entangled
states make to the state space geometry and we have showed it can
solely be attributed to classical correlations for S states and
quantum correlations for W states. Here we have used the relative
entropy of entanglement definition of quantum and classical
correlations and it would be interesting to see what statements can
be made for other entanglement measures. GPs are path dependent
quantities but by determining this path dependence one can learn
something of the geometry of the underlying space. We hope that GPs
may prove useful in determining how classical and quantum
correlations modify this geometry.

MSW \& VV acknowledge funding from EPSRC, QIPIRC and the Royal
Society.

\bibliography{../../../references/masterbib}
\bibliographystyle{apsrev}

\end{document}